\documentclass[superscriptaddress,column,showpacs,a4paper,
amssymb,amsmath,nobibnotes,aps,prd,longbibliography,
showkeys,
nofootinbib,notitlepage]{revtex4-1}
\usepackage{graphicx}
\usepackage{float}
\usepackage{epsf}
\usepackage{bm}
\usepackage{amsmath}
\usepackage{amsfonts}
\usepackage{amssymb}
\usepackage{epstopdf}
\usepackage{natbib}
\usepackage{color}%
\providecommand{\U}[1]{\protect\rule{.1in}{.1in}}

\newcommand{\be}{\begin{equation}}
\newcommand{\ee}{\end{equation}}

\newcommand{\mincir}{\raise
-3.truept\hbox{\rlap{\hbox{$\sim$}}\raise4.truept\hbox{$<$}\ }}
\newcommand{\magcir}{\raise
-3.truept\hbox{\rlap{\hbox{$\sim$}}\raise4.truept\hbox{$>$}\ }}

\newtheorem{remark}{Remark}[section]
\usepackage{xcolor}
\usepackage{hyperref}
\hypersetup{
    colorlinks=true, 
    linkcolor=blue,
    citecolor=magenta}

\begin{document}

\title{
Is Gravity Truly Balanced? A Historical-Critical Journey Through the Equivalence Principle and the Genesis of Spacetime Geometry
}

\author{Jaume de Haro}
\email{jaime.haro@upc.edu}
\affiliation{Departament de Matem\`atiques, Universitat Polit\`ecnica de Catalunya, \\ Diagonal 647, 08028 Barcelona, Spain}

\author{Emilio Elizalde}
\email{elizalde@ice.csic.es}
\affiliation{Institut de Ciències de l'Espai, ICE/CSIC and IEEC, Campus UAB, C/Can Magrans, s/n, 08193 Bellaterra, Barcelona, Spain}

\begin{abstract}
We present a novel derivation of the spacetime metric generated by matter, without invoking Einstein’s field equations. For static sources, the metric arises from a relativistic formulation of D’Alembert’s principle, where the inertial force is treated as a real dynamical entity that exactly compensates gravity. This leads to a conformastatic metric whose geodesic equation---parametrized by proper time---reproduces the relativistic version of Newton’s second law for free fall.
To extend the description to moving matter---uniformly or otherwise---we apply a Lorentz transformation to the static metric. The resulting non-static metric accounts for the motion of the sources and, remarkably, matches the weak-field limit of general relativity as obtained from the linearized Einstein equations in the de Donder (or Lorenz) gauge.
This approach---at least at Solar System scales, where gravitational fields are weak---is grounded in a new dynamical interpretation of the Equivalence Principle. It demonstrates how gravity can emerge from the relativistic structure of inertia, without postulating or solving Einstein’s equations.


\end{abstract}

\vspace{0.5cm}

\keywords{\textit{vis insita}; Principle of Equivalence; Lorentz transformation; approximation to General Relativity}

\maketitle

\section{Introduction}

The so-called Weak Equivalence Principle (WEP) has historically served as conceptual starting point for the formulation of General Relativity (see \cite{Norton} for a historical--philosophical perspective on the Equivalence Principle). Albert Einstein himself confessed on several occasions that the idea he had in 1907 that, if he was falling freely from the roof of his house, he would have eliminated gravity in his immediate surroundings, was what inevitably led him to this profound principle.

{It is worth noting that Ernst Mach may have influenced this idea, as he described the sensation experienced when falling as the cessation of internal pressures caused by weight \cite{Mach}:}
\begin{quote}
    \textit{When we jump or fall from a certain height, we experience a particular sensation that must come from the cessation of the pressures that the particles of the body (from blood, etc.) exert on one another due to their weight.}
\end{quote}

{At its core}, the WEP states that  inertial and gravitational masses are one and the same. And this goes far beyond the also important fact that all bodies fall with the same acceleration, in a gravitational field, regardless of their mass or composition. This last observation, inherited from Galileo and systematized by Newton---\textit{{and understood in terms of force balance by D'Alembert}
}---finds in Einstein a radical extension: if, in a state of free fall, there is no sensation of gravity, then there is no way to distinguish, via local physical experiments, between a uniform gravitational field and a uniformly accelerated reference frame---\textit{{a principle famously summarized as Einstein dixit}}.

However, it is important to emphasize that this formulation has often led to {misunderstandings}, particularly in basic pedagogical contexts. It {is frequently} claimed that “gravity completely disappears” in free fall or that an accelerated system can actually “simulate” the presence of a gravitational field, implying that acceleration would generate a kind of “fictitious gravity”.  This interpretation, stemming from certain classical readings of the Equivalence Principle, has also contributed to the consolidation of the concept of fictitious forces. Things must be set in their proper terms.

{Historically,} the term “fictitious force” began to be used systematically in 19th-century treatises to describe effects like the centrifugal force or the Coriolis force, which arise when analyzing motion from non-inertial reference frames. \textit{{These forces were not presented as real physical interactions}} but rather as \textit{{mathematical constructs}} introduced to preserve the validity of Newton’s second law in accelerated frames. Already in the works of William Thomson (Lord Kelvin) and Edward Routh, these forces were regarded as useful tools but conceptually distinct from real forces, and this view became consolidated in analytical mechanics and terrestrial dynamics texts, towards the end of the 19th century.

This tradition later influenced some interpretations of the Equivalence Principle, where acceleration was associated with a “fictitious gravity” and presumed to be interchangeable with real gravity. However, as we discuss here, strictly speaking, this equivalence \textit{{only makes sense locally}} and within very specific frameworks. It clearly demands a more careful interpretation, from both a physical and a geometric standpoint.

{In contrast,} classical mechanics already contains a deeper concept of the dynamical role of inertia. In the Scholium that opens the \textit{{Principia Mathematica}} \cite{Newton1687}, Newton defines the {{vis insita}} as the ``innate force of matter,'' a natural force residing in every body, and which constitutes its resistance to any change of state. This inertial force is real: it manifests itself whenever a body is forced to deviate from its state of rest or uniform rectilinear motion and gives rise to observable reactions. Thus, Newton did not conceive of inertia as an illusion or a mathematical artifact, but rather as an inherent physical property of bodies.

Although Newton formalized the concept of inertial mass in his second law of motion, the empirical concept that bodies resist changes in velocity---that is, exhibit inertia---was already implicit in Galileo’s experiments and reasoning. Galileo was the first to recognize that all bodies, regardless of composition or weight, fall with the same acceleration in the absence of friction. This is equivalent to having empirically observed that inertial mass $m_\mathrm{i}$---which measures resistance to change in motion---is proportional to gravitational mass $m_\mathrm{g}$---which measures the intensity with which a body experiences gravitational attraction. The equality $m_\mathrm{i} = m_\mathrm{g}$, fundamental to the Equivalence Principle, was thus anticipated experimentally long before its full dynamical formulation in Newton's work, let alone its elevation to the category of a fundamental principle by Einstein.

Most probably, Galileo could not have fully confirmed this in the much-referenced experiment of the Leaning Tower of Pisa (which may never have occurred, in fact), but he did confirm it in detail through several other experiments with tilted planes.

{

Keeping in mind Newton’s vis insita and how D’Alembert later reformulated Newton’s second law as a static balance between real and inertial forces, the purpose of this work is to recover that dynamic conception of inertia and to reinterpret the Equivalence Principle accordingly. In our view, free fall in a gravitational field does not imply the absence of forces, but rather their exact compensation, as in D’Alembert’s principle. This perspective leads naturally to a reformulation of gravity---not as a geometric postulate but as a dynamical consequence of the structure of inertia. By identifying the relativistic inertial force with the spatial component of the four-force in Special Relativity, we construct a spacetime metric that incorporates gravity into the relativistic framework without invoking Einstein’s field equations.

From this standpoint, a freely falling body is not exempt from physical interactions but is instead subject to two real effects, gravitational and inertial, that exactly cancel each other. The Equivalence Principle thus becomes a principle of dynamic compensation rather than one of gravitational suppression. This interpretation does not deny the empirical equivalence between free fall and inertial motion but rather deepens its meaning: instead of treating free fall as the absence of gravity, we understand it as a state in which gravity is precisely balanced by the body’s dynamical response to spacetime geometry.

This reinterpretation helps clarify a conceptual tension already present in the early stages of relativistic theory. When Einstein applied the Equivalence Principle to describe a uniform gravitational field, he derived a metric with temporal curvature but spatial flatness---an indication that the Equivalence Principle alone does not produce full spacetime curvature. The case of free fall aboard the International Space Station makes this evident: astronauts feel weightless due to the precise compensation between gravitational and inertial effects, yet the surrounding spacetime is curved. This illustrates that acceleration alone can mimic some local effects of gravity but cannot reproduce the global geometry of a genuine gravitational field \cite{Eli_book_cos}.

In summary, we propose a view of the Equivalence Principle that consistently respects the real nature of gravitational interaction and recognizes the objective existence of the inertial force as the body’s reaction. Taking this into account, we aim to introduce gravity within a metric theory by modifying the Minkowski metric---not by postulating Einstein’s field equations, but by relying on the physical principles underlying Newtonian dynamics and Lorentz invariance.

Starting from the relativistic form of Newton’s second law, formulated with proper time as the dynamical parameter, and demanding that the geodesic motion derived from a metric reproduce this law, we are naturally led to a static spacetime metric associated with a given gravitational potential. This metric, in turn, allows us to describe the gravitational field generated by static mass distributions.

The key point is that this construction does not require a prior assumption of spacetime curvature: curvature arises as a consequence of enforcing consistency between the Equivalence Principle, the inertial reaction force, and relativistic kinematics. Once the static case is established, we then invoke the Principle of Relativity---specifically, Lorentz transformations---to describe the gravitational field as seen by observers moving relative to the static source. In doing so, we obtain a new metric that accounts for gravity produced by moving matter distributions.

Remarkably, this metric---derived without invoking Einstein’s field equations---coincides, in the linear approximation and when adopting the Lorenz gauge, with the one obtained from General Relativity itself. This agreement confirms the internal consistency of the approach  and demonstrates that the metric structure of weak gravitational fields can in fact be derived from basic physical principles without postulating the full framework of Einstein’s theory.

This not only bridges Newtonian mechanics and relativistic gravity in a conceptually transparent way but also offers a pedagogically effective framework for understanding the emergence of the gravitational metric---first in static contexts, and then in dynamical situations involving moving sources---always rooted in the Equivalence Principle as a dynamic law of balance between real forces.

Throughout, natural units are used ($\hbar = c = k_B = 1$), with $G$ denoting the gravitational constant.

}

\section{Einstein and the Equivalence Principle: How He Actually Derived\\ the Metric}

When Albert Einstein introduced his Equivalence Principle, he did not first start from the powerful formalism of Riemannian geometry (see, e.g., \cite{Stachel1980, Renn2007, Norton1984}). Instead, his reasoning was basically grounded in physical intuition. He considered what an observer would experience inside a spaceship undergoing uniform acceleration in empty space. In his seminal 1911 paper \cite{Einstein1911}, which built upon his earlier exploration of gravitation in 1907 \cite{Einstein1907}, Einstein proposed that no local experiment could ever distinguish between uniform acceleration and a uniform gravitational field.

From this idea, he studied how measurements of time and space would be modified within a system under constant acceleration.

Let us pause for a moment, as the importance of this issue deserves emphasis. The very fact that measurements of time and space are relative---alongside the discovery of the equivalence between mass and energy---was already one of the greatest contributions of his special theory of relativity. It was the materialization of a vision he had at the age of 16, while still a student at the cantonal school in Aarau, Switzerland. Einstein imagined himself chasing a beam of light through the vacuum of space. For several years, he returned to that vision, wondering what would happen if he could actually reach the speed of light \cite{Einstein1951}:

\begin{quote}
\textit{“A paradox I discovered at the age of sixteen: if I pursue a beam of light traveling at the speed $c$, I should observe the beam as a spatially oscillating electromagnetic field at rest. However, it does not seem that such a concept exists. In this paradox already lay the germ of the special theory of relativity. Today we know that any attempt to clarify the paradox satisfactorily was doomed to failure as long as axioms such as the absoluteness of time or of simultaneity remained deeply rooted in our subconscious. Clearly recognizing that we were always using these axioms, and realizing their arbitrary nature, already implied that we had the essential insight needed to resolve the problem. Gradually, I lost hope in discovering the true laws through constructive efforts based on ‘well-established facts.’ The more I tried, the more convinced I became that only the discovery of a universal formal principle could lead to secure results.”} \\
(Einstein, 1951)
\end{quote}

As he acknowledged, the above paradox led him to abandon the absolute nature of time and space (these were, in this case, the ``well-established facts'' he refers to in the quote), which in turn led him to formulate special relativity in 1905. By deriving the Lorentz transformations from the principle of (Galilean) covariance and the constancy of $c$ in vacuum, Einstein brilliantly showed that these expressions were nothing other than the precise mathematical formulation of the fundamental fact that the constancy of the speed of light necessarily implies that time and space must be relative.

This already represented a colossal extension of Galilean relativity, which was later further extended by the Equivalence Principle---though in a different sense, as it was not directly ``absorbed'' into the new theory. It is enough to recall, for example, that in the relativistic corrections applied to GPS measurements, two well-differentiated components appear: one comes from special relativity and the other from general relativity. Both corrections are absolutely essential for the proper functioning of the positioning system.

Returning now to our discussion, in his 1911 paper \cite{Einstein1911}, ``On the Influence of Gravitation on the Propagation of Light'', Einstein developed a remarkable idea. He considered two frames: one, {namely, $K$},  at rest in a homogeneous gravitational field, and the other, {namely, $K'$},  undergoing uniform acceleration in the absence of gravity. He postulated that “these two systems are physically equivalent in all respects,” extending this equivalence beyond the domain of Newtonian mechanics. As he wrote:

\begin{quote}
\textit{As long as we restrict ourselves to purely mechanical processes in the realm where Newton’s mechanics is valid, we are certain of the equivalence of the systems $K$ and $K'$. But our view of this will not have any deeper significance unless the systems $K$ and $K'$ are equivalent with respect to all physical processes.}
\end{quote}

This statement captures the true essence of the Equivalence Principle: “the impossibility of distinguishing, by physical means, between a uniformly accelerated system and one at rest in a homogeneous gravitational field.” Moreover, Einstein writes:

\begin{quote}
\textit{This assumption of exact physical equivalence makes it impossible for us to speak of the absolute acceleration of the system of reference, just as the usual theory of relativity forbids us to talk of the absolute velocity of a system.}
\end{quote}

In other words, acceleration cannot be defined in absolute terms if its physical equivalence with a gravitational field holds. From this, it follows that gravity and acceleration are physical manifestations of the same underlying phenomenon. Consequently, extending the Principle of Relativity to include uniformly accelerated frames naturally leads to the incorporation of gravitational fields---this marks the true essence of the Equivalence Principle.

Following this line of reasoning, Einstein concluded that clocks positioned at higher locations in a gravitational field would tick at a different rate than those located lower. He wrote:

\begin{quote}
\textit{“An observer at rest in the gravitational field experiences that a clock located at a higher altitude runs faster than one located near the ground.”} \\
(Einstein, 1911)
\end{quote}

Based on this difference in clock rates, Einstein deduced that the frequency of a light ray ascending in a gravitational field must decrease. From this effect, he obtained an expression for the variation in frequency with height in a uniform field:
\begin{equation}
\nu = \nu_0 \left(1 + \Phi \right),
\end{equation}
where $\nu_0$ is the frequency emitted at a position with gravitational potential $\Phi = 0$, and $\nu$ is the frequency observed at another position with potential $\Phi$. This relation implies a gravitational time dilation.

Einstein translated this effect into a modification of the spacetime metric in two papers published in 1912 \cite{Einstein1912a, Einstein1912b}. In them, he derived an approximate relationship between an inertial reference frame and a uniformly accelerated one, in the 
$OZ$ direction, expressed through the following transformation:
\begin{eqnarray}
    T = (1 + a z) t, \quad Z = z + (1 + a z) a \frac{t^2}{2}, \quad X = x, \quad Y = y,
\end{eqnarray}
where $a$ is the uniform acceleration.

Applying this transformation to the Minkowski line {element} 
\begin{eqnarray}
ds^2 = dT^2 - d{\bf X} \cdot d{\bf X}
\end{eqnarray}
yields
\begin{eqnarray}
ds^2 = (1 + a z)^2 dt^2 - d{\bf x} \cdot d{\bf x}.
\end{eqnarray}

Therefore, by invoking the Equivalence Principle, Einstein concluded that in a uniform gravitational field, the metric takes the form:
\begin{eqnarray}
    ds^2 = (1 + g z)^2 dt^2 - d{\bf x} \cdot d{\bf x},
\end{eqnarray}
where $g$ is the gravitational acceleration, which  
led him to consider spatially flat metrics of the form:
\begin{eqnarray}\label{flat}
    ds^2 = c^2({\bf x}) dt^2 - d{\bf x} \cdot d{\bf x},
\end{eqnarray}
to describe static gravitational fields.

{
Although Einstein’s 1912 metric does not describe genuine spacetime curvature in the Riemannian sense, it marked a crucial conceptual breakthrough.
Despite its simplicity and provisional status, this early formulation captured a key physical insight: by replacing acceleration with a gravitational field, the Equivalence Principle naturally points toward the geometrization of gravity.
In this sense, even in the absence of full curvature, the 1912 metric already embodies the central idea that gravity manifests through the structure of spacetime, beginning with the way time is measured.}

\subsection{Historical Evolution of the Metric and Curvature in General Relativity}

As we have already {emphasized}, the development of General Relativity was initially guided by the \textit{{Principle of Equivalence}}, formulated by Einstein in 1907, which states that a uniform gravitational field is locally equivalent to an accelerated frame. This insight led Einstein, in his early attempts, to seek a geometric description of gravity with a spatially flat metric, as we have discussed in depth.

Between 1912 and 1913, Einstein collaborated closely with his mathematician friend Marcel Grossmann, who introduced him to the tools of differential geometry---most notably the Ricci tensor, which naturally emerged as a candidate for expressing the gravitational field equations. However, when applying these geometric tools alongside a spatially flat metric in the weak-field limit, Einstein and Grossmann were unable to recover the classical Poisson equation for the gravitational potential
\cite{Grosmann}. This posed a significant conceptual and mathematical challenge.

Although the Zurich notebooks do not preserve a full record of these calculations, several historians \cite{Straumann,Norton1984,Renn2007} suggest that the difficulty stemmed from the assumption of a spatially flat metric, which restricted the theory’s capacity to accurately represent gravity. Consequently, Einstein initially abandoned the Ricci tensor as the basis for his field equations.

It was not until late 1915, after Einstein successfully derived the precise value of Mercury’s perihelion precession, that he fully embraced the need for a general metric, curved in both spatial and temporal components. In his November 1915 communications, Einstein finally presented his complete gravitational field equations \cite{Einstein1915}:
\begin{equation}\label{Ricci}
\mathfrak{Ric} = 8 \pi G \left( \mathfrak{T} - \frac{1}{2} \mathfrak{g} T \right),
\end{equation}
where $\mathfrak{Ric}$ denotes the Ricci tensor describing spacetime curvature, $\mathfrak{T}$ is the energy--momentum tensor, $T$ its trace, and $\mathfrak{g}$ the metric tensor.

In this formulation, gravity is encoded by the full curvature of spacetime, thus overcoming the limitation of the previously assumed spatially flat metric.

\begin{remark}
We emphasize that here, we present Einstein’s gravitational field equations in the original form introduced during his November 1915 lectures at the Prussian Academy of Sciences \cite{Einstein1915}. We intentionally avoid the more familiar modern form:
\begin{equation}
\mathfrak{Ric} - \frac{1}{2} \mathfrak{g} R = 8 \pi G \mathfrak{T},
\end{equation}
where $R$ is the Ricci scalar curvature.

Although this formulation has become standard---mainly because the Bianchi identities guarantee the divergence-free nature of the left-hand side, ensuring local conservation of energy and momentum---it can obscure the original conceptual development.

Indeed, the conservation of energy--momentum is not simply a mathematical consequence of the field equations. On the contrary, it played a central and constructive role in their formulation. As shown in Einstein’s original derivation for a dust-filled universe, the requirement of energy--momentum conservation served as a key guiding principle in determining the structure of the gravitational equations. Rather than being a mere result, this conservation law acted as a constraint shaping the equations themselves.

This historical perspective highlights the deep interplay between physical principles and mathematical formalism in General Relativity. It reveals how Einstein’s profound physical insight---particularly his insistence on compatibility with known conservation laws---was crucial in uncovering the full geometric nature of gravitation.
\end{remark}

It is also important to note that Einstein’s Equation (\ref{Ricci}) is not only covariant under coordinate transformations (passive diffeomorphisms), but also invariant under a broader class of \textit{{active diffeomorphisms}}, transformations that rearrange points of the spacetime manifold rather than merely relabel coordinates.

In particular, the theory is invariant under so-called \textit{{hole diffeomorphisms}}, which act nontrivially outside the support of the energy--momentum tensor but leave it unchanged where matter or energy exists. In other words, these transformations modify the metric in “empty” regions without affecting physically meaningful regions.

This symmetry has profound implications: it means that solutions of Einstein’s equations are not unique in a strong mathematical sense. There exist multiple geometrically distinct metrics (differing pointwise) that are physically indistinguishable, since they represent the same physical situation up to an active diffeomorphism.

For this reason, General Relativity is fundamentally a \textit{{gauge theory}}: the freedom to select different metric representatives related by diffeomorphisms reflects a redundancy in description rather than a physical difference.

This phenomenon has inspired significant philosophical debate, starting with the famous \textit{{hole argument}}. The central question concerns whether spacetime points have ontological status independent of physical events (substantivalism) or should be understood relationally, in terms of relations among events (relationalism). The invariance under active diffeomorphisms strongly supports the latter view, emphasizing that the physically meaningful content lies not in individual points but in the network of physical relations \cite{Stachel2014,Earman-Norton,Iftime-Stachel,Weatherall,Haro2024}.

Finally, it is worth recalling that simultaneously with Einstein’s definitive formulation of General Relativity in November 1915, the mathematician David Hilbert presented an axiomatic formulation of the gravitational theory based on a variational principle, which has since become widely used. The question of priority and the relationship between these works has been the subject of intense historiographical discussion.

Recent scholarship based on original manuscripts and proofs indicates that Hilbert’s initial submission, shortly before Einstein’s publication, did not contain the field equations in their final form. It appears that Hilbert incorporated the complete equations only after learning of Einstein’s results \cite{Corry}.

This evidence suggests that while Hilbert contributed an elegant and mathematically grounded variational formulation, the priority for the physical insight and the correct final form of the field equations indisputably belongs to Einstein. Historical records also reveal that the intellectual exchange and competition between these two great minds significantly advanced the theory.

This interpretation is now widely accepted by historians such as Jürgen Renn, Michel Janssen, and others who have extensively studied the archival materials of both authors.

\subsection{The True Meaning of the Equivalence Principle According to Einstein in 1911: A Critical Reading}

We have already shown above that one of the most powerful and revolutionary ideas in the history of modern physics is the  Equivalence Principle. However, its standard interpretation---as presented in most textbooks on General Relativity---often diverges significantly from Einstein’s original insights. Here, we revisit Einstein’s 1911 formulation of the principle \cite{Einstein1911}, where he first established a concrete link between gravity and acceleration, grounded in the equivalence of inertial and gravitational mass.

\begin{enumerate}
\item \textbf{{The Standard Interpretation: Free Fall as an Inertial System}}

According to the commonly accepted interpretation, a body in free fall does not feel its own weight. Because of this, it is said that the body can be considered “at rest” in a local inertial frame. Based on this idea, many argue that the Principle of Relativity from Special Relativity---originally valid only in the absence of gravity---can be applied locally to freely falling systems, even in the presence of a gravitational field. This leads to the widespread claim that “gravity disappears” in free fall, and that General Relativity extends Special Relativity by patching together such locally inertial frames.

However, this way of expressing things can be conceptually misleading. The statement that “gravity disappears” sounds more like a philosophical metaphor than a physical explanation. It does not clarify the actual mechanism behind the motion of a falling body, nor does it address the real forces that are at play. As a result, this interpretation fails to capture the true dynamics that the body experiences during free fall.

\item \textbf{{A Weak Gravitational Field and Spatially Flat Metric}}

In this context, as we {have already} explained, Einstein derived a first approximation of the gravitational metric in which only the temporal component was modified, while the spatial part remained flat. At this stage (as already stated before), Einstein did not yet consider gravity as implying spatial curvature. This led him, as we previously noted, to initially reject the Ricci tensor as the foundation for the field equations, since it necessarily entails curvature of the full spacetime geometry. Only later did he come to accept that a correct description of gravity requires the full curvature of spacetime---including both temporal and spatial components.

Einstein’s early rejection of the Ricci tensor clearly shows that, at the beginning of his search for a gravitational theory, he did not yet view gravity as a purely geometric phenomenon. Instead, he still thought of gravity as a physical effect, closely connected to the experience of acceleration. This idea was grounded in his insight that a uniformly accelerated frame is physically equivalent to a homogeneous gravitational field---a key feature of his Equivalence Principle. At that stage, his focus was more on understanding gravity in terms of the forces felt by observers, rather than on describing it through the curvature of spacetime.

In later decades, many authors criticized this early version of the Equivalence Principle, pointing out its ambiguities and limitations. Among the most notable critiques is that of John L. Synge, who openly questioned the physical validity of the principle in the mature framework of General Relativity. In his own words \cite{Synge}:

\begin{quote}
\textit{
I have never been able to understand this Principle of Equivalence. Does it mean that the effects of a gravitational field are indistinguishable from the effects of an observer’s acceleration? If so, it is false. In Einstein’s theory, either there is a gravitational field or there is none, according as the Riemann tensor does not or does vanish. This is an absolute property; it has nothing to do with any observer’s world-line. Space-time is either flat or curved, and in several places in the book I have been at considerable pains to separate truly gravitational effects due to curvature of space-time from those due to curvature of the observer’s world-line\dots The Principle of Equivalence performed the essential office of midwife at the birth of General Relativity\dots I suggest that the midwife be now buried with appropriate honors and the facts of absolute space-time faced.
}
\end{quote}

This provocative statement underscores a deeper conceptual shift: once General Relativity was established as a theory of spacetime curvature, the heuristic role of the Equivalence Principle---essential during the theory’s gestation---actually became less central, and even potentially misleading if interpreted as a literal identity between acceleration and gravitation.

\end{enumerate}

\section{The Inertial Force as a Dynamical Reaction: {A Reinterpretation of the Equivalence Principle}}

We {begin this section with} the following paragraph, written by Newton in his 
\textit{{Principia}} \cite{Newton1687}:

\begin{quote}
\begin{flushleft}
\textit{``The vis insita, or innate force of matter, is a resistance power by which every body, as much as it lies, continues in its present state, whether it be of rest or of moving uniformly forward in a straight line. This force is always proportional to the body whose force it is and differs nothing from the inactivity of the mass, but in our way of conceiving it.''}
\end{flushleft}
\end{quote}

In this passage, Newton introduces a very fundamental idea that every body spontaneously resists any attempt to change its state of motion. This resistance is what he calls \textit{{innate force}} or {{vis insita}}, and it is directly linked to the quantity of matter---what we now call mass.

When Newton says that this force “differs nothing from the inactivity of the mass, except in our manner of conceiving it,” he is emphasizing that it is, fundamentally, the same physical property: inertia.

\begin{enumerate}
    \item If we consider the body at rest or in uniform motion, we see it as passivity: the body simply “maintains” its state.
    \item But if we try to accelerate it, that same property manifests itself as a resisting force: an opposition to change, proportional to the inertial mass, which we denote by $m_\mathrm{i}$, i.e., a force of the type $\mathbf{F}_\mathrm{i} = -m_\mathrm{i} \mathbf{a}$, where $\mathbf{a}$ is the acceleration applied to the body.
\end{enumerate}

Let us now examine how this Newtonian view connects with the Weak Equivalence Principle.

The conventional reading of this principle tends to present free fall as a situation without real forces acting, as if gravity “disappeared” in the absence of a reaction on a scale. However, this idea ignores a richer and more coherent physical interpretation, already present, in germ, in Newtonian mechanics.

Further, in the Scholium that accompanies the initial definitions of the \textit{{Philosophi\ae{} Naturalis Principia Mathematica}}, Newton introduces the concept of {{vis insita}}, which he defines as:

\begin{quote}
\begin{flushleft}
\textit{Vis insita est potentia resistendi, qua corpus unumquodque conatur perseverare in statu suo quiescendi vel movendi uniformiter in directum.}

(The innate force is the power of resisting, by which each body strives to maintain its state of rest or of uniform motion in a straight line.)
\end{flushleft}
\end{quote}

It is worth highlighting that Newton understood inertial mass as a measure of resistance to acceleration but did not equate it with an active force. For him, it was merely a “dynamic passivity.” In contrast, one century later, Jean le Rond d’Alembert introduced a key innovation in his \textit{{Traité de dynamique}} (1743): he transformed Newton’s second law into a dynamic equilibrium equation by adding the so-called “inertial force” to the system. His principle---known today as D’Alembert’s Principle---states \cite{Dalembert1743}:

\begin{quote}
\begin{flushleft}
\textit{To reduce dynamic problems to simple static problems, one may consider that inertial forces $-m \mathbf{a}$ act on the body, opposing the applied forces.}
\end{flushleft}
\end{quote}

And he added:

\begin{quote}
\begin{flushleft}
\textit{I call, as Mr. Newton does, the force of inertia the property that bodies have of remaining in the state in which they are: it is this property that we must demonstrate here.}
\end{flushleft}
\end{quote}

We thus see how D’Alembert emphasizes that inertia is not simply the absence of action but a real (albeit internal) force that resists change of state, in line with Newton’s idea of {{vis insita}}. With this vision, he formulates his principle of dynamic equilibrium:

\begin{quote}
\begin{flushleft}
\textit{In a system, the internal inertial forces are equal and opposite to the forces that produce the acceleration.}

(D’Alembert 1757)
\end{flushleft}
\end{quote}

From this viewpoint, the {{vis insita}}---also called {{vis inertiae}}---is not simply another name for inertia, but an ontological assertion: the resistance to changing motion is a natural force inherent to the body. D’Alembert considers it a real force, observable when a body is made to accelerate or decelerate, and proportional to the inertial mass:
$$
\mathbf{vis\ insita} = -m_\mathrm{i} \mathbf{a},
$$
where $\mathbf{a}$ is the acceleration. This conception is central to Newton’s second law, which establishes that the force applied to a body is proportional to its acceleration:
$$
\mathbf{F} = m_\mathrm{i} \mathbf{a},
$$
and can be viewed as a force balance $\mathbf{F} + \mathbf{F}_\mathrm{i} = 0$: the inertial force compensates the applied force \cite{Wenceslao}.

From this perspective, the inertial force is not a mathematical construct but a real dynamic reaction of the body---a measurable manifestation of its vis insita. For example, consider the outward push felt in a curve (centrifugal force), or when a train brakes suddenly and passengers are thrown forward: what manifests itself is the persistence of their state of motion. The body reacts to the deceleration imposed by the train. This reaction is no illusion or fiction: it can be measured, recorded, and causes real physical effects (such as injuries in the absence of seat belts).

In the gravitational context, we have the following. The force pulling a body toward the ground is
$$
\mathbf{F}_\mathrm{g} = m_\mathrm{g} \mathbf{g},
$$
where $m_\mathrm{g}$ is the gravitational mass. This generates an inertial force
$$
\mathbf{F}_\mathrm{i} = -m_\mathrm{i} \mathbf{a},
$$
which compensates the gravitational force. And, as Galileo already observed in experiments, all bodies fall with the same acceleration $\mathbf{a} = \mathbf{g}$, which implies that the values of inertial and gravitational mass are proportional. We can then choose a system of units such that this proportionality constant equals one, leading to the numerical equality:
$$
m_\mathrm{i} = m_\mathrm{g}.
$$

Thus, the situation of free fall does not involve an absence of forces, but rather a dynamic equilibrium between two real and opposite forces: gravitational attraction and the inertial reaction of the body. This approach is consistent with the Newtonian conception of dynamics and, as we show in the next section, extends naturally to the framework of General Relativity.

According to the present interpretation, this is the true physical content of the Weak Equivalence Principle:
In a local freely falling system, the acceleration of the body generates an inertial force that exactly compensates the gravitational force. Since, according to Einstein, the inertial and the gravitational masses of any object are equal,
$$
m_\mathrm{i} = m_\mathrm{g},
$$
it follows that the gravitational force acting on a body is proportional to its resistance to acceleration. As a result, all objects---regardless of their mass or composition---experience the same acceleration when falling freely in a gravitational field.

When one reinterprets the Equivalence Principle as a condition of dynamic equilibrium between two real forces---following D’Alembert’s approach---one gets a much clearer and more tangible understanding. Instead of relying on the abstract idea that gravity and acceleration are locally “indistinguishable,” this viewpoint emphasizes a physical balance: the inertial force exactly cancels the gravitational force during free fall.

{

\subsection{{The} 
 Modified Inertial Law on Galactic Scales}

 It is well known that in spiral galaxies, stars orbit the galactic center in nearly circular trajectories. According to Newton's law of gravitation, the centripetal force required to maintain such motion must be provided by the gravitational pull of the mass enclosed within the star's orbit. Assuming a spherically symmetric mass distribution, this implies that the orbital velocity \( v(r) \) of a star located at a distance \( r \) from the center should decrease with \( r \) as
\begin{eqnarray}
v(r) \propto \frac{1}{\sqrt{r}},
\end{eqnarray}
once the bulk of the mass lies within the radius \( r \).

However, astronomical observations tell a different story. Far beyond the visible edge of the galactic disk---where little luminous matter is detected---the rotational velocity of stars does not decrease as predicted but rather stays approximately constant. These so-called \emph{{flat rotation curves}} are in clear tension with the expectations based on visible matter alone.

To preserve the validity of Newtonian gravity in this regime, it is commonly postulated that galaxies are embedded in a halo of invisible matter---known as \emph{{dark matter}}---that provides the additional gravitational pull needed to maintain the observed velocities at large distances.

An alternative explanation, originally proposed by Milgrom in the early 1980s and known as Modified Newtonian Dynamics (MOND) \cite{Milgrom}, challenges this assumption. Instead of invoking unseen mass, MOND proposes a modification of Newton's second law of motion in the regime of extremely small accelerations. Specifically, Milgrom suggested that below a critical acceleration scale \( a_0 \sim H_0 \) (where \( H_0 \) is the present value of the Hubble rate, and we use natural units with \( c = 1 \)), the inertial response of a body is no longer linear. The standard expression \( \mathbf{F}_{\rm i} = -m \mathbf{a} \) is replaced by
\begin{equation}
\mathbf{F}_{\rm i} = -m \, \mu\left( \frac{|\mathbf{a}|}{a_0} \right) \mathbf{a},
\end{equation}
where \( \mu(x) \to 1 \) when \( x \gg 1 \), recovering Newtonian dynamics at high accelerations, and \( \mu(x) \to x \) when \( x \ll 1 \), introducing a significant deviation in the low-acceleration regime. Remarkably, this modification naturally leads to flat rotation curves, without requiring the presence of dark matter.

A notable aspect of this approach is that the gravitational field remains Newtonian; it is instead the law of inertia that is altered. This conceptual shift is particularly interesting in light of approaches that reexamine the inertial structure of spacetime itself.

This viewpoint resonates with the perspective developed in the present work, where the inertial force is also regarded as a real, relativistic effect that dynamically balances the gravitational interaction. Although the theoretical foundations differ, both approaches share the core idea that inertia may not be an immutable law, but rather a dynamical response that could require revision under certain physical conditions.

However, embedding this modified law of inertia within a relativistic framework remains highly challenging, as illustrated in \cite{Milgrom1}, where the difficulties in constructing a classical action that incorporates this inertial force are thoroughly discussed.
One possible route, which we adopt in what follows, is inspired by the work of Bekenstein and Milgrom \cite{Bekenstein}. There, the modification of inertia is effectively reinterpreted as a modification of the gravitational potential.

Let \( \Phi \) be the standard Newtonian potential, and let \( \bar{\Phi} \) denote the modified potential. The relation between both is given by
\begin{equation}
\mu\left( \frac{|\nabla \bar{\Phi}|}{a_0} \right) \nabla \bar{\Phi} = \nabla \Phi.
\end{equation}
{Taking} 
 the divergence of both sides and applying the classical Poisson equation, we obtain the modified Poisson equation:
\begin{equation}
\nabla \cdot \left[ \mu\left( \frac{|\nabla \bar{\Phi}|}{a_0} \right) \nabla \bar{\Phi} \right] = 4\pi G \rho,
\end{equation}
where \( \rho \) is the matter density.

Consequently, all results derived in the following sections using the Newtonian potential \( \Phi \) remain valid when \( \Phi \) is replaced by the MONDian potential \( \bar{\Phi} \). This correspondence allows us to incorporate the phenomenology of modified inertia into our relativistic framework in a consistent and effective manner.

}

\section{Incorporation of Non-Flat Spatial Metric from Geodesics}

In this section, we see how the view defended here of the Equivalence Principle leads us to a spatially non-flat metric.
The goal is to incorporate gravity in the context of special relativity, which suggests simply replacing the Minkowski interval by another one that includes the gravitational potential, and whose geodesics satisfy equations as close as possible to Newton's second law, so as to identify the equivalence of gravity and the inertial force.

Building on D’Alembert’s perspective, we start with the idea that the motion of a freely falling body---when observed from a rest frame in a homogeneous and static gravitational field---follows Newton’s second law. We then introduce a key physical assumption: that the body moves along a path that extremizes its proper time, just as light follows the path that minimizes travel time in Fermat’s principle. This assumption naturally leads us to the geodesic equation, the cornerstone of motion in curved spacetime.

Using this principle, we can derive the form of the spacetime metric that corresponds to the given gravitational field. Moreover, this method can be generalized, at least within the weak-field approximation, to determine the metric associated with a wider class of static gravitational fields.

To build intuition about the appropriate metric structure, we apply the Equivalence Principle to a uniform gravitational field described by the potential  
\begin{equation}
    \Phi(z) = gz,
\end{equation}
and require Newton’s second law to hold exactly:
\begin{equation}\label{inertia}
    \mathbf{F}_{\rm i} + \mathbf{F}_{\rm g} = 0,
\end{equation}
where the inertial force is given by  
\begin{equation}
    \mathbf{F}_{\rm i} = -m \mathbf{a}_{\rm p} \equiv -m (0,0,a_{\rm p}),
\end{equation}
and the gravitational force is  
\begin{equation}
    \mathbf{F}_{\rm g} = -m (0,0,g).
\end{equation}

To derive the corresponding dynamical equation from a relativistic perspective, we assume that freely falling bodies follow geodesics, i.e., trajectories that extremize the proper time. This is conceptually analogous to Fermat’s principle in optics, where light follows the path that extremizes travel time.

Assuming spatial isotropy, we propose a general static line element of the form:
\begin{equation}\label{uniform}
    ds^2 = A(z) \, dt^2 - B(z) \, d\mathbf{x} \cdot d\mathbf{x}.
\end{equation}
{From} the principle of stationary action,
\begin{equation}
    \delta S = \delta \int ds = 0,
\end{equation}
we obtain Newton’s second law in the form of Equation~\eqref{inertia}, provided we identify:
\begin{equation}\label{exact}
    a_{\rm p} = \frac{d^2 z}{ds^2}, \quad A(z) = 1 + 2 \Phi(z), \quad B(z) = \frac{1}{1 + 2 \Phi(z)}.
\end{equation}

We conclude that, for a uniform gravitational field $\Phi(z) = gz$, a freely falling body satisfies Newton’s second law when its proper acceleration $a_{\rm p}$ corresponds to the spatial component of the relativistic four-acceleration defined in Special Relativity, and the spacetime metric is no longer spatially flat. That is, the gravitational field manifests itself as a genuine curvature of spacetime.

Moreover, this interpretation provides a concrete explanation for why a body in free fall does not feel its own weight: the gravitational force is exactly compensated by the inertial reaction force, understood here as the spatial projection of the four-force in Special Relativity. This stands in contrast to the more ambiguous and commonly stated idea that ``gravity disappears'' in free fall. Gravity does not disappear; rather, it is precisely balanced by a real, dynamical inertial force.

Therefore, based on the  intuition given by the study of a uniform gravitational field, it is natural to look for a still simple metric whose line element is \textit{{conformastatic}} (i.e., static and isotropic):
\begin{eqnarray}\label{conformastatic_metric}
ds^2 = (1 + 2 \Phi({\bf x})) dt^2 - \frac{1}{1 + 2 \Phi({\bf x})} d{\bf x} \cdot d{\bf x}.
\end{eqnarray}

This line element was initially proposed by J. Synge \cite{Synge} and used to describe static gravitational fields.
The geodesic equation associated with this metric is:
\begin{eqnarray}\label{eq_geodesica}
\ddot{\bf x} = - \nabla \Phi - 2 \frac{|\dot{\bf x}|^2}{1 + 2 \Phi} \nabla^\perp \Phi\cong 
- \nabla \Phi - 2 {|\dot{\bf x}|^2} \nabla^\perp \Phi,
\end{eqnarray}
where the derivative is with respect to proper time, and we have introduced the component of the gradient of the potential perpendicular to the velocity:
\begin{eqnarray}
\nabla^\perp \Phi = \nabla \Phi - \frac{\dot{\bf x} \cdot \nabla \Phi}{|\dot{\bf x}|^2} \dot{\bf x}.
\end{eqnarray}

When the velocity is parallel to the gradient of $\Phi$, the exact Newtonian equation is recovered:
\begin{eqnarray}
\ddot{\bf x} = - \nabla \Phi,
\end{eqnarray}
which reflects the \textit{{Weak Equivalence Principle}} if the inertial force (the vis insita) acting on a particle is interpreted as the spacial component of the  four-force, namely,
\begin{eqnarray}
{\bf F}_{\rm i} = - m \ddot{\bf x},
\end{eqnarray}
and the gravitational force  is
\begin{eqnarray}
{\bf F}_{\rm g} = - m \nabla \Phi.
\end{eqnarray}

Thus, the dynamical equation becomes
\begin{eqnarray}
{\bf F}_{\rm i} + {\bf F}_{\rm g} = 0 \quad \Longrightarrow \quad - m \ddot{\bf x} - m \nabla \Phi = 0.
\end{eqnarray}

\begin{remark}
    Note the difference between our metric approach and the non-metric one proposed by Max {Abraham} 
 \cite{Abraham}, where the author postulated the following relativistic equations:
    \begin{eqnarray}
        \ddot{\mathbf{x}} = -\nabla \Phi \qquad \text{and} \qquad
        \ddot{t} = \partial_t \Phi,
    \end{eqnarray}
where the potential $\Phi$ is a scalar field because it satisfies the relativistic generalization of the Poisson equation:
\begin{eqnarray}\label{D'Alambert_eq}
    \Box \Phi=-4\pi G\rho,
\end{eqnarray}
where $\Box=\partial_{t^2}^2-\Delta$ is the D'Alambertian operator in flat spacetime, and $\rho$ is the rest density of matter, and thus, a scalar.

    For a static potential, this implies \( \dot{t} = 1 \). In contrast, our metric approach, based on the line element (\ref{conformastatic_metric}), leads to
    \begin{eqnarray}
    \dot{t}^2 = \frac{1}{1 + 2\Phi} \left( 1 + \frac{|\dot{\mathbf{x}}|^2}{1 + 2\Phi} \right) \cong 1 + |\dot{\mathbf{x}}|^2,
    \end{eqnarray}
    in the weak-field approximation.

    Moreover, in Abraham's framework, the tidal force
    \begin{eqnarray}\label{tidal}
        \mathbf{F}_{\rm tidal} \equiv -2m
        \frac{|\dot{\mathbf{x}}|^2}{1 + 2\Phi} \nabla^{\perp} \Phi
    \end{eqnarray}
    does not appear.
\end{remark}

In particular, for a freely falling body near the Earth's surface following a radial trajectory, the equation of motion reduces to
\begin{eqnarray}
\ddot{r} = - \frac{G M}{r^2} \approx 9.8\, \mathrm{m/s}^2,
\end{eqnarray}
demonstrating the exact compensation between inertial and gravitational forces, which is nothing but the manifestation of Einstein's Equivalence Principle in its original formulation.

Moreover, the conformastatic metric leads to energy conservation, since multiplying the geodesic Equation (\ref{eq_geodesica}) by the momentum $m \dot{\bf x}$ yields
\begin{eqnarray}
\frac{m |\dot{\bf x}|^2}{2} + m \Phi({\bf x}) = E,
\end{eqnarray}
where $E$ is a constant energy.

Once the physical reasons to choose this static metric are justified, in the weak field approximation, we have
\begin{eqnarray}
\frac{1}{1 + 2 \Phi} \approx 1 - 2 \Phi,
\end{eqnarray}
and the line element becomes
\begin{eqnarray}\label{METRIC}
ds^2 = (1 + 2 \Phi) dt^2 - (1 - 2 \Phi) d{\bf x} \cdot d{\bf x}.
\end{eqnarray}

This metric naturally incorporates non-flat spatial curvature and is consistent with the predictions of General Relativity in the weak gravity regime. 
For example, the metric correctly predicts the perihelion precession of Mercury \cite{Haro2025}, a phenomenon that puzzled astronomers for decades. 
In its orbit around the Sun, the point closest to the star---the perihelion---does not remain fixed but slowly rotates over time. Newton's laws, together with gravitational perturbations from other planets, explained almost all of this precession but not completely: there was a small discrepancy of about 43 arcsecond per century without apparent explanation. 
Einstein \cite{Einstein1915a}, applying his General Relativity theory up to the second order in perturbations, calculated the correction due to the curvature of spacetime caused by the Sun and obtained exactly this missing value. This result was one of the first great successes of the theory and provided crucial evidence of its validity.

{
Another striking consequence of Einstein’s General Theory of Relativity---and, of course, of the metric (\ref{METRIC})---is the bending of light as it passes near a massive object such as the Sun. Although Einstein had initially estimated this deflection using a preliminary and less complete version of his theory, it was only after deriving the full Schwarzschild solution in 1916---following the formulation of his field equations---or alternatively using the metric (\ref{METRIC}), that the correct value of the bending angle was obtained. This remarkable prediction was famously confirmed by Arthur Eddington during the 1919 solar eclipse, when he measured the apparent displacement of stars near the edge of the Sun---an effect caused by the curvature of spacetime produced by the Sun’s mass.

}

In fact, the full calculation becomes much more elegant using the Schwarzschild metric, which Einstein did not have when he first tackled the problem. Nonetheless, once he derived the correct deflection angle, Einstein became fully convinced of the soundness of his theory---long before it was confirmed by observation. The success of Eddington’s expedition propelled Einstein to global fame as “the physicist who had defeated Newton.”

When a journalist asked Einstein what he would have thought had the eclipse results failed to match his predictions, he reportedly replied: “Then I would have been sorry for the good Lord---the theory is correct.”

\section{Incorporation of Motion via Lorentz Transformations}

Following \textit{{Kenneth Nordtvedt}}~\cite{Nordtvedt1968,Will},  consider a reference frame \( K' \) moving with velocity \( -\mathbf{v}' \) relative to a frame \( K \). Equivalently, \( K \) moves with velocity \( \mathbf{v}' \) with respect to \( K' \).

Let \( \gamma \) be the Lorentz factor:
\begin{equation}
\gamma = \frac{1}{\sqrt{1 - |\mathbf{v}'|^2}}.
\end{equation}

The infinitesimal Lorentz transformation relating the coordinates in both frames is:
\begin{equation}\label{eq:LorentzInfinitesimal}
    d\mathbf{x} = \gamma \left[
    \frac{1}{\gamma} \, d\mathbf{x}' + \frac{\gamma}{1 + \gamma} (\mathbf{v}' \cdot d\mathbf{x}') \mathbf{v}' - \mathbf{v}' \, dt'
    \right], \quad
    dt = \gamma \left( dt' - \mathbf{v}' \cdot d\mathbf{x}' \right).
\end{equation}

Now, by dropping the primes and keeping only linear terms in the velocity \( \mathbf{v}' \), the line element---originally conformastatic in the frame \( K \)---takes the form
(see for more details \cite{Haro2025a})
:
\begin{equation}\label{eq:MetricMovimiento}
    ds^2 = (1 + 2\Phi) \, dt^2 -8\, \mathbf{N} \cdot d\mathbf{x} \, dt - (1 - 2\Phi) \, d\mathbf{x} \cdot d\mathbf{x},
\end{equation}
where  the vector field 
${\bf N}=\Phi {\bf v} $,
arising from the Lorentz transformation, encodes the gravitomagnetic components induced by the motion.

Note that this line element coincides with the metric obtained in the linearized General Relativity approximation in the (harmonic) de Donder gauge   \cite{Fock}.

To deduce these potentials in the case of dust-like matter ($p \ll \rho$), it is crucial to recognize the cumulative nature of the gravitational effect: each infinitesimal element of the moving mass distribution contributes independently to the total metric. Since these elements may have different velocities, a Lorentz transformation must be applied locally to each of them, according to their particular velocities. The global potential is then obtained by summing---more precisely, by integrating---all these contributions over the entire mass distribution, ensuring that the resulting metric properly reflects the system's dynamic configuration.

The relativistic generalization of the Newtonian gravitational potential, as proposed by M. Abraham~\cite{Abraham} (see also~\cite{Nordstrom}), must satisfy the relativistic Poisson Equation (\ref{D'Alambert_eq}):
\begin{equation}\label{Phi}
    \Box \Phi = -4\pi G \rho.
\end{equation}

In parallel, by analogy with the relativistic formulation of the Biot--Savart law, the gravitational vector potential \( \mathbf{N} \) must satisfy the condition:
\begin{equation}\label{Ene}
    \Box \mathbf{N} = -4\pi G \rho \, \mathbf{v},
\end{equation}
where the source term corresponds to the mass current density.

{
\begin{remark}
Retaining quadratic terms in the velocity, the line element acquires the form:
\begin{eqnarray}
ds^2 &=& (1 + 2\Phi + 4\Upsilon) \, dt^2 - 8\, \mathbf{N} \cdot d\mathbf{x} \, dt \nonumber\\
&& - (1 - 2\Phi + 4\Phi^2) \, d\mathbf{x} \cdot d\mathbf{x} + 4\, \mathfrak{h}_{ij} \, dx^i dx^j,
\end{eqnarray}
where the new potentials $\Upsilon$ and $\mathfrak{h}_{ij}$ satisfy the equations:
\begin{eqnarray}
\Box \Upsilon = -4\pi G \rho |\mathbf{v}|^2, \qquad \Box \mathfrak{h}_{ij} = -4\pi G \rho v_i v_j,
\end{eqnarray}
which also coincide with the result obtained in linear General Relativity in harmonic gauge when one retains quadratic terms in the velocity (see \cite{Einstein1970} and Section 110 of \cite{Landau}).
\end{remark}
}

\begin{remark}
    By analogy with electrodynamics, we can define the gravitational four-potential 
    ${\bf A} = (\Phi, \mathbf{N})$ and the four-current 
    ${\bf J} = (\rho, \rho \mathbf{v})$, so that the field equation for the potentials takes the compact form:
    \begin{equation}\label{XXX}
        \Box {\bf A} = -4\pi G {\bf J}.
    \end{equation}

Then, 
the classical continuity equation 
\begin{eqnarray}
\partial_t \rho + \nabla \cdot (\rho \mathbf{v}) = 0 \quad \Longleftrightarrow \quad \mbox{div}({\bf J}) = 0,
\end{eqnarray}
corresponds to the so-called \textit{Lorenz gauge}
\begin{eqnarray}
\partial_t \Phi + \nabla \cdot \mathbf{N} = 0 \quad \Longleftrightarrow \quad \mbox{div}({\bf A}) = 0.
\end{eqnarray}

{\bf {Historical note:}}
The name Lorenz gauge refers to the Danish physicist Ludvig Valentin Lorenz, who introduced this condition in 1867. However, it is often mistakenly called the Lorentz gauge due to the similarity of names with the Dutch physicist Hendrik Antoon Lorentz, known for the Lorentz transformations in special relativity. Although this distribution is widespread, it is historically inaccurate. See, for instance, (\cite{Jackson}, p.~663, abstract) for a historical clarification.

On the other hand, 
one can also introduce the \textit{gravitoelectromagnetic fields}:
\begin{eqnarray}
    {\bf E} \equiv -\nabla \Phi - \partial_t {\bf N}, \qquad \text{and} \qquad {\bf H} \equiv \nabla \wedge {\bf N},
\end{eqnarray}
which, based on their definitions and Equation (\ref{XXX}), satisfy the following gravitoelectromagnetic equations:
\begin{eqnarray}
    \nabla \cdot {\bf E} = -4\pi G \rho, \qquad
    \nabla \cdot {\bf H} = 0, \qquad
    \nabla \wedge {\bf E} + \partial_t {\bf H} = 0, \qquad
    \nabla \wedge {\bf H} - \partial_t {\bf E} = -4\pi G \rho {\bf v},
\end{eqnarray}
which are the gravitational analog of Maxwell's equations.

\end{remark}

These field equations are complemented by the classical conservation law and Euler's equation. However, to remain closer in spirit to General Relativity, we impose the conservation of the energy--momentum tensor, expressed as \( \mathrm{div}(\mathfrak{T}) = 0 \). For example, when dealing with a perfect fluid, where the energy--momentum tensor takes the form:
\begin{equation}
    \mathfrak{T} = (\rho + p) \, \mathfrak{u} \otimes \mathfrak{u} - p \, \mathfrak{g},
\qquad \mbox{with}\qquad p\ll \rho, 
\end{equation}
where {$\frak{g}$ denotes the tensor metric and}
 \( \mathfrak{u} \) is the one-form dual to the four-velocity \( \mathbf{u}  \), {  that is, $\frak{u}={\bf u}^{\flat}$ 
where the \emph{{flat operator}} is defined below}.
This conservation law leads to the following evolution equations for a relativistic fluid:
\begin{equation}\label{conservationeq}
    \frac{{ d} \rho}{ds} = -(\rho + p) \, \mathrm{div}(\mathbf{u}), \qquad
    (\rho + p) \, \frac{D \mathbf{u}}{ds} = \mathrm{grad}(p) - \frac{{d} p}{ds} \, \mathbf{u},
\end{equation}
where \( D/ds \) denotes the covariant derivative along its worldline, and the gradient of the pressure is defined as the vector associated with the differential one-form \( dp \) via the metric, that is, using  the so-called \emph{{sharp operator}} ({Given} 
 a Riemannian or pseudo-Riemannian manifold with metric $g$, the \emph{{sharp operator}} \( \sharp \) maps a covector (one-form) \( \omega = \omega_\mu dx^\mu \) to the corresponding vector field \( \omega^\sharp = g^{\mu\nu} \omega_\mu \partial_\nu \). It is defined as the inverse of the musical isomorphism \( \flat \), which maps a vector \( v = v^\mu \partial_\mu \) to the one-form \( v^\flat = g_{\mu\nu} v^\nu dx^\mu \). These operations are fundamental in formulating differential geometry in coordinate-free language.):
{\begin{eqnarray}
    \mathrm{grad}(p)\equiv (dp)^{\sharp}.
         \end{eqnarray}}
         
\noindent and the divergence of the four-velocity
{\bf u}
is the trace  of the endomorphism that maps any 
 four-vector ${\bf w}$ to  the covariant derivative $\nabla_{\bf w}{\bf u}$:
\begin{eqnarray}
    \mathrm{div}(\mathbf{u})\equiv \mbox{Tr}\left({\bf w}\rightarrow
    \nabla_{\bf w}{\bf u}\right).\end{eqnarray}

{

In addition, taking into account that 
$$
\frac{dp}{ds} = \mathfrak{g}({\rm grad}(p), \mathbf{u}),
$$
we can use the \emph{{orthogonal rejection}} of the gradient of the pressure onto $\mathbf{u}$, namely,
$$
{\rm grad}^{\perp}(p) \equiv {\rm grad}(p) - \mathfrak{g}({\rm grad}(p), \mathbf{u}) \, \mathbf{u},
$$
to write the Equation~(\ref{conservationeq}) as follows:
\begin{equation}\label{conservationeq1}
    \frac{d \rho}{ds} = -(\rho + p) \, \mathrm{div}(\mathbf{u}), \qquad
    \frac{D \mathbf{u}}{ds} = \frac{1}{\rho + p} \, {\rm grad}^{\perp}(p).
\end{equation}

Finally, note the well-known fact that for a dust fluid, i.e., a pressureless fluid, the flow follows the geodesics of spacetime:
\begin{equation}\label{conservationeq2}
    \frac{d \rho}{ds} = -\rho \, \mathrm{div}(\mathbf{u}), \qquad
    \frac{D \mathbf{u}}{ds} = 0.
\end{equation}


}


\subsection{Geodesic Equation}

To derive the geodesic equations to first order in perturbations, we recall that the potential $\Phi$ is of the order $\frac{MG}{L} \ll 1$, where $L$ is the characteristic length scale of the motion. 
Since $|\dot{\bf x}|^2$ is of the same order as $\Phi$ (from energy conservation), we can assume that the main velocity of the objects, ${\bf v}$, is of the order $\sqrt{\frac{MG}{L}}$, which is always satisfied at the scale of the Solar System. Therefore, the vector potential ${\bf N} = \Phi {\bf v}$ is of the order $\left(\frac{MG}{L}\right)^{3/2}$.

Moreover, the gradient operator $\nabla$ is of order $\frac{1}{L}$, and the time derivative $\partial_t \sim \dot{\bf x} \cdot \nabla$ is of order $\frac{1}{L} \sqrt{\frac{MG}{L}}$. 
Finally, $\ddot{\bf x}$ is of order $\frac{MG}{L^2}$, and since $\dot{t} \sim 1 - \Phi + \frac{|\dot{\bf x}|^2}{2}$, it follows that $\ddot{t}$ is of order $\left(\frac{MG}{L}\right)^{3/2} \frac{1}{L}$.

Hence, up to order $\frac{M^2 G^2}{L^3}$, for the line element (\ref{eq:MetricMovimiento}), the geodesic equation is approximately (for details, see \cite{Haro2025a}):
\begin{eqnarray}\label{Eq_G}
   \ddot{\bf x}&=&-\nabla\Phi-4\partial_{ t}{\bf N}+4 \dot{\bf x}\wedge (\nabla\wedge {\bf N})+2\partial_t\Phi\dot{\bf x}
    -{2}|\dot{\bf x}|^2\nabla^{\perp}\Phi
    \nonumber\\
    &=&4({\bf E}+\dot{\bf x}\wedge{\bf H})+
\left(2\partial_t\Phi\dot{\bf x}+3
\nabla^{\parallel}\Phi
    \right)
    +(3-{2}|\dot{\bf x}|^2)\nabla^{\perp}\Phi    ,
    \end{eqnarray}
where $\nabla^{\parallel}\Phi$ is the orthogonal projection of the gradient of $\Phi$ onto the direction of $\dot{\bf x}$, and the first term on the right-hand side is four times the gravitational analog of the Lorentz force.

Effectively, the Lagrangian coming from the metric (\ref{eq:MetricMovimiento}) is:
\begin{eqnarray}
    \mathcal{L} = (1 + 2\Phi)\dot{t}^2 - 8\,
    \dot{t}{\bf N} \cdot \dot{\bf x}  - (1 - 2\Phi)|\dot{\bf x}|^2 \qquad \text{with} \qquad \mathcal{L} = 1.
\end{eqnarray}

Then, up to order $\frac{M^2 G^2}{L^3}$, we have
\begin{eqnarray}
    \frac{d}{ds}\left( 
    \partial_{\dot{\bf x}} \mathcal{L} \right)
    = -2 \left[
        4(\partial_t {\bf N} + (\dot{\bf x} \cdot \nabla){\bf N}) 
        - 2(\partial_t \Phi + \dot{\bf x} \cdot \nabla \Phi)\dot{\bf x} 
        + (1 - 2\Phi)\ddot{\bf x}
    \right],
\end{eqnarray}
and
\begin{eqnarray}
    \partial_{\bf x} \mathcal{L} =
    2 \nabla \Phi (1 - 2\Phi + 2|\dot{\bf x}|^2) 
    - 8 \left[ (\dot{\bf x} \cdot \nabla){\bf N} + \dot{\bf x} \wedge \nabla \wedge {\bf N} \right].
\end{eqnarray}

Therefore, the Euler--Lagrange equation leads to (\ref{Eq_G}).

In particular, if we consider the field created by a homogeneous sphere of radius $R$ rotating with constant angular velocity ${{\bf w}}$, we get
\begin{eqnarray}
   { \ddot{\bf x}=-\frac{MG}{|{\bf x}|^3}{\bf x}-2{\bf \Omega}\wedge\dot{\bf x}
    -2\frac{MG|\dot{\bf x}|^2}{|{\bf x}|^3}{\bf x}_{\perp}},
\end{eqnarray}
where ${\bf \Omega}=\frac{IG}{|{\bf x}|^5}(|{\bf x}|^2{{\bf w}}-3({\bf x}\cdot {{\bf w}}){\bf x})$ with $I=\frac{2MR^2}{5}$ being the moment of inertia, and ${\bf x}_{\perp}$ the component of ${\bf x}$ orthogonal to the velocity.

{

\subsection{The Cauchy Problem}

At this point, dealing with a pressureless fluid---which includes planets---we are able to formulate our Equations (\ref{Phi}), (\ref{Ene}) and (\ref{conservationeq2}) as an evolution problem with initial conditions. Since we are considering the weak-field limit, we retain only the leading-order terms of the equations.

First of all, from relation (\ref{eq:MetricMovimiento}),
\begin{equation}
\frac{ds^2}{dt^2} = (1+2\Phi) - 8{\bf N} \cdot {\bf v} - (1-2\Phi){\bf v} \cdot {\bf v},
\end{equation}
where ${\bf v} = \frac{d{\bf x}}{dt}$, we find
$$
\frac{ds}{dt} \cong \sqrt{1+2\Phi} \cong 1 + \Phi,
$$
and thus,
$$
\frac{d{\bf x}}{ds} = \frac{dt}{ds} {\bf v} \cong (1 - \Phi){\bf v} \cong {\bf v}.
$$

Therefore, dealing with the geodesic equation $\frac{D{\bf u}}{ds} = 0$, and taking into account that in our approximation, ${\bf u} \cong (1 - \Phi, {\bf v})$, and retaining the leading term of (\ref{Eq_G}), one finds
\begin{equation}
\frac{d{\bf v}}{ds} = -\nabla \Phi \quad \Longrightarrow \quad
\frac{d{\bf v}}{dt} = -\nabla \Phi \quad \Longrightarrow \quad
\partial_t {\bf v} = -\nabla \Phi,
\end{equation}
where we have used that $\frac{d{\bf v}}{dt} = \partial_t {\bf v} + ({\bf v} \cdot \nabla){\bf v} \cong \partial_t {\bf v}$.

Next, we deal with the conservation equation $\frac{d\rho}{ds} = -\rho\, {\rm div}({\bf u})$. Using $\sqrt{-g} \cong 1 - 2\Phi$, we find
$$
{\rm div}({\bf u}) \cong -3\partial_t \Phi + \nabla \cdot {\bf v},
$$
and thus, the leading term of the conservation equation is
\begin{equation}
\partial_t \rho = -\nabla \cdot (\rho {\bf v}) + 3\rho\, \partial_t \Phi.
\end{equation}

Then, the three dynamical equations are:
\begin{equation}\label{P1}
\Box \Phi = -4\pi G \rho, \qquad
\partial_t \rho = -\nabla \cdot (\rho {\bf v}) + 3\rho\, \partial_t \Phi, \qquad
\partial_t {\bf v} = -\nabla \Phi,
\end{equation}
with initial conditions at $t = t_0$:
\begin{equation}\label{C1}
\Phi(t_0, {\bf x}) = \Phi_0({\bf x}), \qquad
\partial_t \Phi(t_0, {\bf x}) = \Psi_0({\bf x}), \qquad
\rho(t_0, {\bf x}) = \rho_0({\bf x}), \qquad
{\bf v}(t_0, {\bf x}) = {\bf v}_0({\bf x}).
\end{equation}

Note also that once the problem (\ref{P1}) has been solved numerically---since it is impossible to solve it analytically---with the initial conditions (\ref{C1}), we obtain the values, at all points of spacetime, of the Newtonian potential, the mass density, and the velocity field. In addition, since we know the values of $\rho$ and ${\bf v}$, we can determine ${\bf N}$ by solving the equation $\Box {\bf N} = -4\pi G \rho {\bf v}$ using the method of retarded potentials. Therefore, we also obtain the full line element, which allows us to determine the motion of test particles.

In summary, our method reveals that we can, at least for weak fields, transform Einstein's field equations into a simple Cauchy problem without resorting to the complicated ADM formalism \cite{ADM}.

}

\section{Physical Interpretation and Reformulation of the Equivalence Principle}

{In this final section, we clarify the core idea behind our reformulation of the Equivalence Principle and emphasize the key differences between our approach and Einstein’s original interpretation.}

Starting with  Newton's second law:
\begin{equation}
\frac{d^2 {\bf x}}{dt^2} = - \nabla \Phi,
\end{equation}
which, after identifying gravitational and inertial mass, and reasoning in the spirit of D'Alembert, can be rewritten as
\begin{equation}
- m \frac{d^2 {\bf x}}{dt^2} - m \nabla \Phi = 0 \quad \Longrightarrow \quad {\bf F}_{\rm i} + {\bf F}_{\rm g} = 0.
\end{equation}

That is, the inertial force ${\bf F}_{\rm i}$ and the gravitational force ${\bf F}_{\rm g}$ exactly cancel out, so the body does not feel its weight.  
Therefore, D'Alembert’s principle, by introducing the inertial force as a real entity that compensates external forces, anticipates---purely from a mechanical point of view---the  idea that gravity can be locally canceled by acceleration. While D'Alembert did not formulate this concept in terms of subjective experience or spacetime geometry, his dynamical principle lays a foundation that is conceptually compatible with the modern reinterpretation of free fall that led to General Relativity.

{On the other hand, as we have already shown, D'Alembert's principle can be extended to a metric theory---particularly within the relativistic framework---by replacing the classical inertial force with the \textit{{spatial component of the four-force}}, which represents the actual force acting on the body. Accordingly, the geodesic equation for a conformastatic metric can be expressed as:}
\begin{equation}
{\bf F}_{\rm i} + {\bf F}_{\rm g} + {\bf F}_{{\rm tidal}} = 0,
\end{equation}
where the \textit{{tidal force}} is defined in (\ref{tidal}). 
{Moreover, when the velocity vector is parallel to the gradient of the potential, the equation precisely reduces to Newton’s second law.}

\begin{remark}
In General Relativity, the motion of a freely falling body is governed by the \textit{geodesic equation}, written in terms of the covariant derivative:
\begin{equation}
    \frac{D {\bf u}}{ds} = 0,
\end{equation}
where 
\( \mathbf{u} = \left( \frac{dt}{ds}, \frac{d{\bf x}}{ds} \right) = \left( \frac{dt}{ds}, \mathbf{v} \right) \)
is the four-velocity of the particle. This equation states that the four-velocity is \textit{parallel-transported along itself}, meaning that the particle follows a geodesic, i.e., the straightest possible path in curved spacetime.

The notion of \textit{four-acceleration} generalizes its special relativistic counterpart. Instead of using the flat-spacetime derivative \( d{\bf u}/ds \), one defines the \textit{geometric four-acceleration} as:
\begin{equation}
    {{\bf a}_{\rm f}\equiv} \frac{D {\bf u}}{ds}.
\end{equation}

Thus, a freely falling body has zero geometric four-acceleration, since no real force acts on it: its motion is entirely dictated by the geometry of spacetime. Consequently, the \textit{four-force}---defined as the product of mass and four-acceleration---also vanishes:
\begin{equation}
    {\bf f} = m \, {{\bf a}_{\rm f}} = 0.
\end{equation}

This expresses the condition for free motion in General Relativity: the absence of any real force acting on the particle. Even though an external observer may describe the particle's trajectory as accelerated, the particle itself experiences no proper acceleration---it feels no resistance---and simply follows the path determined by spacetime curvature.

In contrast, our interpretation is fundamentally dynamical, inspired by D’Alembert’s principle, where the inertial force is regarded as a real and measurable quantity. Within Special Relativity, we identify the inertial force with the spatial component of the four-force. In this approach, spacetime geometry does not determine motion {a priori}; instead, the metric structure emerges from the requirement that, when the particle's velocity is parallel to the gradient of the gravitational potential, the inertial and gravitational forces exactly cancel out during free fall.

Thus, rather than interpreting geodesic motion as a purely geometric feature of spacetime, we view it as the result of an exact balance of physical forces. The vanishing of the four-force---expressed via the covariant derivative---is not postulated but derived from the dynamical condition of compensation between inertia and gravity.
\end{remark}

Let us now analyze the main differences between this reformulation and Einstein’s original version of the Equivalence Principle:
\begin{enumerate}
    \item \textbf{{Interpretation of forces}}: In Einstein’s early formulation, the Equivalence Principle is usually stated as the local indistinguishability between gravitational fields and acceleration. That is, gravity can be ``transformed away'' by choosing a freely falling frame. In our interpretation---\textit{{closer to D'Alembert's than  to Einstein’s }}---both gravitational and inertial forces are real and coexist, and they cancel out exactly when
    $m_{\rm i} = m_{\rm g}.$

    The inertial force is defined as the spatial component of the four-force in Special Relativity:
    $$
    {\bf F}_{\rm i} = - m_{\rm i} {\bf a}_{{\rm p}},
    $$
    which exactly balances the gravitational attraction when the motion is along the gradient of the potential.

    \item \textbf{{Origin of the inertial force}}: In this framework, the inertial force acting on a freely falling body is not fictitious. Rather, it is a real dynamical reaction experienced by the body, similar to the force felt when accelerating in a vehicle. The sensation of weightlessness arises from this exact compensation, not from the disappearance of gravity.

    \item \textbf{{Metric interpretation}}: While Einstein initially viewed acceleration as generating an equivalent gravitational field---illustrated by his famous elevator thought experiment---our approach shows that imposing the exact cancellation of gravitational and inertial forces during free fall, together with the principle of least action (i.e., minimization of proper time), leads---at least in the weak-field limit---to the metric associated with a static gravitational field.

    \item \textbf{{Critique of field generation by acceleration}}: Our approach explicitly rejects the notion, present in Einstein’s early writings, that acceleration itself generates a gravitational field. Instead, we assert that only mass--energy curves spacetime; acceleration alone does not produce curvature. Thus, the equivalence between acceleration and gravity must be understood dynamically, in terms of compensating forces, and not as a statement about field generation.
\end{enumerate}

We hope this reformulation helps clarify the physical content of the Equivalence Principle and facilitates the consistent inclusion of metrics with non-flat spatial curvature, such as the conformastatic metric discussed earlier.

\textbf{{A final comment:}} {The methodology presented in this work represents the culmination of our entire reasoning. It clearly demonstrates that, by locally applying Lorentz transformations and then accumulating their effects across a moving mass distribution, one can generalize the static conformastatic metric to include sources in motion. The resulting spacetime structure coincides with that obtained from the linearized approximation of General Relativity in harmonic gauge, thereby confirming the validity of the present approach.}


This inevitably leads to a puzzling thought: actually, Einstein already had at his disposal both the necessary ideas and mathematical methods, so why did he not take this simpler alternative path?

By the time he formulated Special Relativity, Einstein had a solid grasp of Lorentz transformations and understood that space and time were relative concepts. He also realized that Newton’s notion of gravity acting instantaneously across distances conflicted with the relativistic framework. Then, one might reasonably assume that by combining the principles of relativistic motion with classical gravity, he could have derived an approximate gravitational metric valid for weak fields---and all this without needing to rethink spacetime geometry from the very ground up. Yet, curiously, this was not the route he followed.

As we have discussed above, the reason seems to lie in Einstein’s unique interpretation of the Equivalence Principle. By recognizing a profound equivalence between a uniformly accelerated frame and a homogeneous gravitational field, he concluded that acceleration could no longer be regarded as absolute. This insight led him to treat both gravitational fields and accelerated systems within a relativistic framework, effectively extending the scope of the Principle of Relativity. However, this approach initially led him to envision spacetime without spatial curvature, which introduced significant difficulties in recovering classical mechanics as the limiting case of his new gravitational theory.

Consequently, rather than pursuing a relativistic reformulation of Newtonian gravity, Einstein took a new, more ambitious path: he redefined gravity not as a force, but as a manifestation of the geometry of spacetime. This conceptual leap ultimately led him to the Einstein field equations, which relate the curvature of spacetime (expressed through the Ricci tensor) to the distribution of matter and energy (encoded in the stress--energy tensor). We should here add a further reason: the approach he chose was also much closer to Mach's ideas, which Einstein tried to reflect in his theory \cite{EBSC}.

In summary, Einstein chose a more far-reaching and visionary path, one that was not just an extension of existing theories but a profound reconceptualization of fundamental physics. Even though by 1907, he had at his disposal many of the elements needed to anticipate some important aspects of his final theory, he opted instead for a radical unification: the idea that gravity and inertia are not separate forces but expressions of the same underlying geometric structure, the curvature of spacetime.

\section{Conclusions}

In this work, we developed a framework to derive the gravitational metric starting from the Equivalence Principle and a relativistic version of Newton’s second law, using proper time as the natural parameter. By incorporating Lorentz transformations, we extended the static gravitational metric to scenarios where the matter distribution moved, obtaining a metric consistent with the weak-field approximation of Einstein’s equations in the Lorenz gauge.

A key novelty of our approach is its dynamic interpretation of the Equivalence Principle, inspired by D'Alembert's principle, where inertial forces are real physical reactions balancing gravitational attraction when inertial and gravitational masses coincide. This clarifies why a freely falling body experiences no proper force, providing a more intuitive understanding of the Equivalence Principle.

We emphasize that acceleration alone does not generate gravitational fields; the true source of spacetime curvature is the distribution of mass--energy. This distinction helps separate inertial dynamics from geometric effects and avoids common misconceptions.

Overall, our method bridges classical mechanics and relativistic gravity without solving Einstein’s field equations, offering a pedagogically transparent and physically grounded derivation of the gravitational metric in weak fields. This contributes both to the conceptual foundations and to practical understanding of gravitational phenomena in regimes relevant to the Solar System and similar contexts.

\

\begin{acknowledgments}

JdH is supported by the Spanish grant PID2021-123903NB-I00
funded by MCIN/AEI/10.13039/501100011033 and by ``ERDF A way of making Europe''. EE is partly supported by the program Unidad de Excelencia
María de Maeztu, CEX2020-001058-M, and by AGAUR, project 2021-SGR-00171.  
\end{acknowledgments}

\end{document}